\documentclass{article}
\usepackage{spconf,amsmath,graphicx}
\usepackage[super]{nth}
\usepackage{xurl}
\usepackage[hidelinks,breaklinks=true]{hyperref}
\usepackage[symbol]{footmisc}
\usepackage{siunitx}
\usepackage{comment}
\usepackage{newtxmath,newtxtt}

\usepackage{enumitem}
\usepackage{cleveref}
\Crefname{figure}{Fig.}{Figs.}
\setlist{nosep, leftmargin=14pt}
\usepackage{nicefrac,upgreek,stmaryrd}
\usepackage{booktabs}
\usepackage{multirow}
\usepackage{tablefootnote,threeparttable,array}
\usepackage[dvipsnames]{xcolor}
\usepackage{fancybox}
\definecolor{MyBlue}{HTML}{3b76af}
\definecolor{MyOrange}{HTML}{ef8636}

\title{Cross-Age and Cross-Site Domain Shift Impacts on Deep Learning-Based White Matter Fiber Estimation in Newborn and Baby Brains} 

\name{%
\begin{tabular}{@{}c@{}}%
Rizhong Lin$^{1,2,3}$\qquad%
Ali Gholipour$^{4}$\qquad%
Jean-Philippe Thiran$^{1,2,5}$\qquad%
Davood Karimi$^{4}$\\%
Hamza Kebiri$^{2,5,\star}$\qquad%
Meritxell Bach Cuadra$^{5,2,\star}$%
\thanks{
$^{\star}$ H. Kebiri and M. Bach Cuadra --- Equal contribution. 
}
\end{tabular}
}
\address{%
{%
\resizebox{0.98\linewidth}{!}{%
\begin{tabular}{@{}c@{}}%
$^{1}$ Signal Processing Laboratory 5 (LTS5), \'Ecole Polytechnique F\'ed\'erale de Lausanne (EPFL), Lausanne, Switzerland\\
$^{2}$ Department of Radiology, Lausanne University Hospital (CHUV) and University of Lausanne (UNIL), Lausanne, Switzerland\\
$^{3}$ College of Electronic and Information Engineering, Tongji University, Shanghai, China\\
$^{4}$ Computational Radiology Laboratory, Department of Radiology, Boston Children’s Hospital and Harvard Medical School, Boston, MA, USA\\
$^{5}$ CIBM Center for Biomedical Imaging, Switzerland%
\end{tabular}%
}}%
}%

\usepackage{tikz}
\usepackage{atbegshi}
\usepackage{lipsum} 

\usepackage[
    backend=biber,
    style=ieee,
    natbib=true,
    maxnames=2,
    minnames=1,
    doi=false,
    isbn=false,
    url=false,
    eprint=false,
    date=year,
    giveninits=true,
]{biblatex}
\defbibheading{bibliography}[\bibname]{%
\section{#1}%
\markboth{#1}{#1}}
\DeclareUnicodeCharacter{202F}{!!!!FIX ME!!!!}
\AtEveryBibitem{%
  \clearlist{language}%
}
\begin{document}
%
\maketitle
\begin{abstract}%
Deep learning models have shown great promise in estimating tissue microstructure from limited diffusion magnetic resonance imaging data. However, these models face domain shift challenges when test and train data are from different scanners and protocols, or when the models are applied to data with inherent variations such as the developing brains of infants and children scanned at various ages. Several techniques have been proposed to address some of these challenges, such as data harmonization or domain adaptation in the adult brain. However, those techniques remain unexplored for the estimation of fiber orientation distribution functions in the rapidly developing brains of infants. In this work, we extensively investigate the age effect and domain shift within and across two different cohorts of 201 newborns and 165 babies using the Method of Moments and fine-tuning strategies. Our results show that reduced variations in the microstructural development of babies in comparison to newborns directly impact the deep learning models' cross-age performance. We also demonstrate that a small number of target domain samples can significantly mitigate domain shift problems. 
\end{abstract}
\begin{keywords}
Diffusion MRI, Fiber Orientation Distribution estimation, white matter, domain shift, deep learning
\end{keywords}
\section{Introduction}
\label{sec:intro}

The human brain undergoes notable changes during development, particularly in white matter tracts that modulate cognitive and motor functions \cite{dubois_early_2014}. Accurately estimating these fibers is crucial for understanding developmental patterns and detecting abnormalities. 
Advances in diffusion magnetic resonance imaging (dMRI) have provided unprecedented insights into the human brain microstructure. Traditional methods, such as Constrained Spherical Deconvolution (CSD) \cite{tournier_robust_2007} and Multi-Shell Multi-Tissue Constrained Spherical Deconvolution (MSMT-CSD) \cite{jeurissen_multi-tissue_2014}, have been employed to reconstruct fiber orientation distribution functions (FODs) as a proxy to the underlying microstructure. These methods often require a large number of diffusion measurements and/or multiple $b$ values, making them less feasible for uncooperative young subjects. Recently, deep learning (DL) on large datasets has allowed precise FOD estimation \cite{karimi_learning_2021, hosseini_cttrack_2022, yao_robust_2024} with as few as six diffusion samples from developing brains \cite{kebiri_robust_2023}.

While DL can offer significant scanning time reduction, it is particularly faced with domain-shift problems. Such shifts can be attributed to several factors, from biological differences \cite{bento_deep_2022} such as age or pathologies \cite{dubois_early_2014} to imaging variations in protocols and scanner types (brand or field strength) \cite{tax_cross-scanner_2019}. 

Data harmonization has been used for reducing variability across sites while preserving data integrity \cite{pinto_harmonization_2020}. The dominant dMRI method operating at the signal level is the Rotation Invariant Spherical Harmonics (RISH) \cite{cetin_karayumak_retrospective_2019} that harmonizes dMRI data without model dependency but requires similar acquisition protocols and site-matched healthy controls.
Deep learning techniques offer solutions to non-linear harmonization but risk overfitting and require extensive training data, potentially altering pathological information \cite{bashyam_deep_2022}. The Method of Moments (MoM) \cite{huynh_multi-site_2019}, which aligns diffusion-weighted imaging (DWI) features via spherical moments, stands out for its directionality preservation and independence from matched acquisition protocol or extensive training. Therefore, MoM presents a potentially beneficial approach for addressing the domain shift challenges.

Furthermore, domain adaptation (DA) methods have been used to address domain shifts in medical imaging \cite{guan_domain_2022}. 
Supervised DA, particularly fine-tuning (FT) models with pre-trained weights on source domain data, is a common method, often augmented with advanced, more targeted techniques \cite{zhou_fine-tuning_2017}. Semi-supervised DA methods, which leverage a mix of labeled and unlabeled data, can also effectively bridge domain shifts. However, both semi-supervised and unsupervised DAs face challenges in the case of significant anatomical differences, such as those between infants and neonates, where the assumption of feature space similarity may not hold.

This paper investigates the domain shift effects in a DL method \cite{kebiri_robust_2023} for white matter FOD estimation in the newborn and baby populations. Our goal is to provide a detailed examination of the challenges associated with domain shifts, particularly age-related variations between these young cohorts. We propose possible solutions and emphasize the need for robust frameworks that can cope with the unique variability present in the developing brain.

\section{Methodology}
\label{sec:method}

\subsection{Data Processing}

We used dMRI data from the \nth{3} release of the Developing Human Connectome Project (dHCP) \cite{hutter_time-efficient_2018} and the Baby Connectome Project (BCP) \cite{howell_uncumn_2019}. The dHCP dataset includes 783 subjects from 20--44 post-menstrual weeks, acquired using a 3T Philips scanner and a multi-shell sequence ($b \in$ \{0, 400, 1000, 2600\}~s/mm\textsuperscript{2}). Its dMRI data release has been preprocessed by SHARD \cite{christiaens_scattered_2021}, with a final data resolution of 1.5\textsuperscript{3}~mm\textsuperscript{3}. \textit{White Matter} and \textit{Brainstem} masks, which dHCP also provides, were registered to the dMRI data space and combined with voxels of fractional anisotropy (FA) $>$ 0.3 to produce the final white matter (WM) mask. The BCP dataset comprises 285 subjects from 0--5 years, scanned using a 3T Siemens scanner with a different multi-shell protocol ($b \in$ \{500, 1000, 1500, 2000, 2500, 3000\}~s/mm\textsuperscript{2}). Denoising and bias, motion, and distortion corrections \cite{andersson_integrated_2016} also yielded a 1.5\textsuperscript{3}~mm\textsuperscript{3} resolution for the BCP dMRI data. The final WM mask was established using an OR operation among STAPLE \cite{warfield_simultaneous_2004}-generated WM mask, FA $>$ 0.4, and voxels with FA $>$ 0.15 and mean diffusivity (MD) $>$ 0.0011.
We also computed the mean FA value within the white matter mask of each subject to analyze the relationship between age and FA value.

\subsection{Model}

\begin{figure}[t!]
    \centering
    \includegraphics[width=\linewidth]{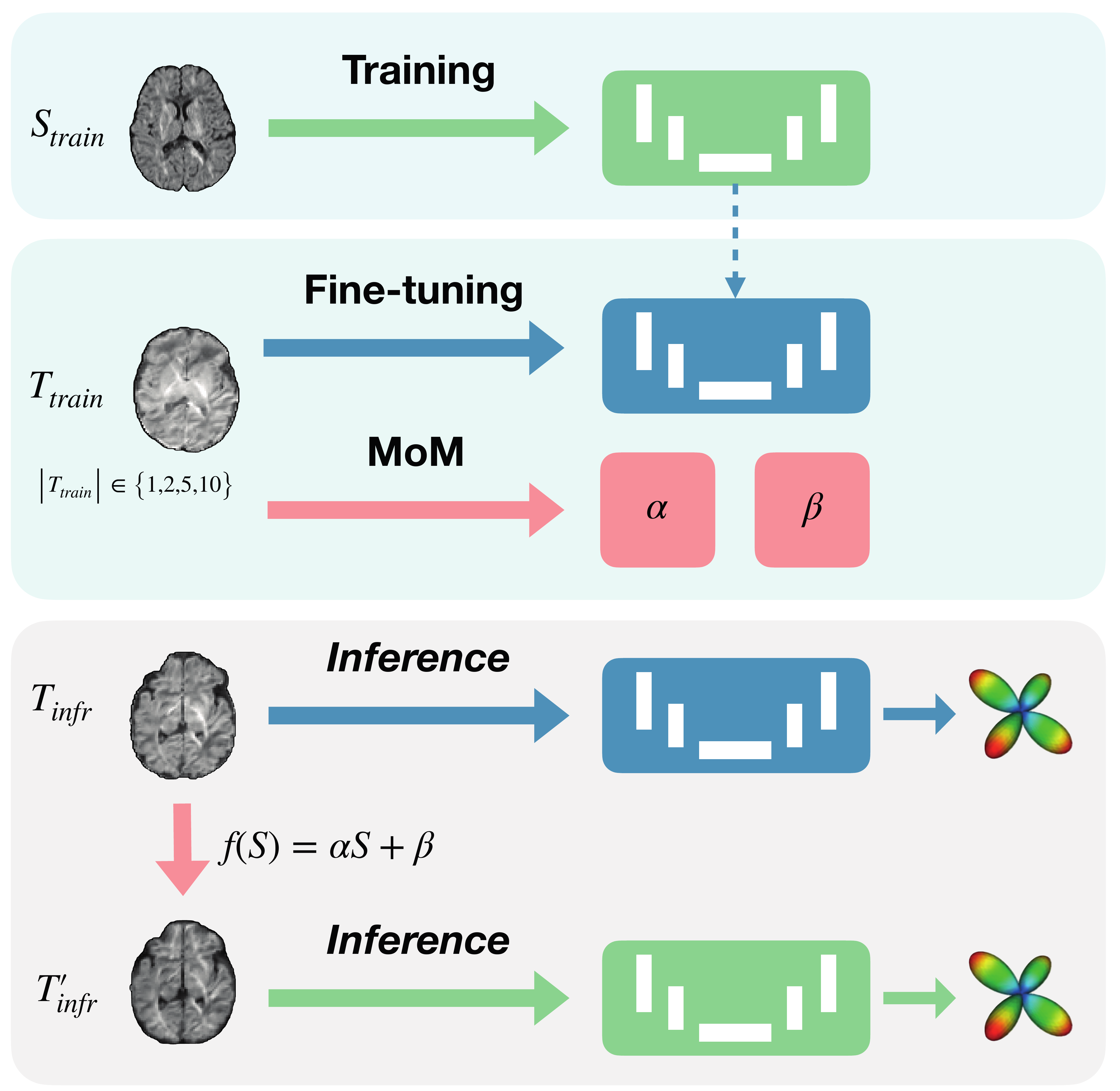}
\caption{
Diagram of the workflow separated into: 
(1) initial model training on the source dataset \(S_{\text{train}}\), (2) MoM harmonization and model fine-tuning applied independently on the target training dataset \(T_{\text{train}}\)  with varying subject numbers (\{1,2,5,10\}), and (3) inference 
where the original model assesses harmonized data \(T'_{\text{infr}}\), and the fine-tuned model evaluates the original target data \(T_{\text{infr}}\).
}
    \label{fig:framework}
\end{figure}

As illustrated in \Cref{fig:framework}, our model's workflow is divided into three stages: initial training on a source dataset, followed by separate processes of either fine-tuning or MoM harmonization on the target dataset with varying number of subjects. During inference, harmonized data is tested using the originally trained model to evaluate the effectiveness of MoM, while the fine-tuned model is applied to the original target data to assess the improvements in FOD estimation.

\subsubsection{Backbone Model}

We employed the U-Net-like network \cite{kebiri_robust_2023} 
as the backbone for our experiments. Its proficiency lies in estimating accurate FODs from dMRI data with six diffusion directions with an extensive field of view (FoV) of $16\times 16\times 16$, and its demonstrated accurate results when applied to dHCP newborns. 

We applied the MSMT-CSD \cite{jeurissen_multi-tissue_2014} using all measurements (i.e., 300 diffusion directions for dHCP and 151 directions for BCP) to generate ground-truth (GT) FODs for training and evaluation.
To ensure a representative sample, subjects were randomly selected from the datasets based on the desired age range and number of subjects required by each experiment detailed in \Cref{sec:intra-site,sec:inter-site}.
For each subject, we processed the diffusion signal by selecting six optimal gradient directions, normalizing, projecting onto the spherical harmonic (SH)-basis, and cropping to $16^3$ patches as in \cite{kebiri_robust_2023}.

For each experiment, we trained the backbone network for 1000 epochs using the Adam optimizer \cite{kingma_adam_2015} with an initial learning rate of \num{5e-5}, weight decay of \num{1e-3}, and batches of 35. We used a dropout of 0.1 to prevent overfitting. Model selection was based on the lowest mean squared error (MSE) between predicted and GT FODs in the validation set.

\subsubsection{Methods for Addressing Domain Shifts}

We explore two primary data harmonization and domain adaptation strategies to handle domain shifts. 

\textbf{Data Harmonization using Method of Moments}
\label{sec:data_harmonization}

MoM \cite{huynh_multi-site_2019} was employed to harmonize DWI data across sites by aligning the mean and variance using linear mapping functions \( f_{\theta=\{\alpha,\beta\}}(S) = \alpha S + \beta \). 
This approach adjusts each voxel's DWI signal \( S \) to the reference site's characteristics. Median images of these moments, smoothed with a Gaussian filter to mitigate artifacts, were computed from the six optimal gradient directions and used to derive the harmonization parameters \( \alpha \) and \( \beta \).

\textbf{Domain Adaptation using Fine-Tuning} 
\label{sec:fine-tuning}

This process involved knowledge transfer and additional training of the model on the target domain data. 
We conducted fine-tuning  over 100 epochs, 
with a reduced learning rate of \num{5e-6} and smaller batches of 10.

\subsubsection{Implementation Details and Code Availability}

Training and fine-tuning were performed on an NVIDIA RTX 3090 GPU. We use TensorFlow 2.11 for our DL framework and MATLAB R2022b for MoM harmonization. Our code is publicly available at \url{http://github.com/Medical-Image-Analysis-Laboratory/dl_fiber_domain_shift}.

\subsection{Intra-Site Age-Related Evaluation}
\label{sec:intra-site}

We assessed baseline performance on the dHCP and BCP datasets, selecting 100 subjects from specified age ranges (dHCP: 29.3--44.3 post-menstrual weeks; BCP: 1.5--60 postnatal months), and allocated them into training, validation, and testing sets (70/15/15). The backbone model was trained and tested on these splits, respectively for BCP and dHCP. GT consistency was evaluated by processing two mutually exclusive subsets of the full measurements with MSMT-CSD (referred to as Gold Standards, GS), as in \cite{kebiri_robust_2023}.

To investigate age-related shifts within each site, we conducted age-specific training and cross-testing. The dHCP dataset was split into two age groups: [26.7,~35.0] and [40.0,~44.4] weeks, and the BCP dataset into [0.5,~11] and [20,~36] months, denoted as \textit{young} and \textit{old}, respectively. Each group consisted of 60 subjects, split into 40/10/10 partitions for training, validation, and testing, respectively. 
Fine-tuning was also performed across different age groups within dHCP using 5 subjects from the corresponding target age group.

\subsection{Inter-Site Experiments}
\label{sec:inter-site}

To address both age-related and cross-site domain shifts, we conducted cross-testing between dHCP and BCP with respective baseline models from \Cref{sec:intra-site}. 
We also evaluated how varying subject numbers in the target training dataset (1, 2, 5, and 10 subjects) affect the performance of MoM harmonization and fine-tuning. Furthermore, an ablation study involved training a model from scratch on 10 target dataset subjects to verify performance gains beyond target set familiarity.

\subsection{Evaluation Metrics}

We quantitatively assessed FOD estimation accuracy using metrics as per \cite{kebiri_robust_2023}: \textit{Agreement Rate (AR)} for peak count consistency, \textit{Angular Error (AE)} for angular discrepancy between predicted and GT FODs, and \textit{Apparent Fiber Density (AFD)} from \cite{raffelt_apparent_2012} to evaluate fiber density.

\section{Results}
\label{sec:results}

\subsection{Intra-Site Experiments}


The DL metrics are first compared to the GT consistency (GS in \Cref{tab:bcp-dhcp-baseline}) for dHCP and BCP. As previously reported in \cite{schilling_histological_2018}, single-fiber predictions show good agreement, but performance decreases with multiple fibers for both datasets. This is more pronounced in three-fiber cases, which exhibit low DL performance as also reported in \cite{kebiri_robust_2023}. We therefore not consider 3-fiber metrics in subsequent experiments.

\begin{table}[t!]
\centering
\caption{Results of baseline DL models and GS of dHCP and BCP. AR and AE are reported for 1-, 2-, and 3-fiber (1/2/3-F) configurations, alongside AFD Error ($\Delta$AFD).}
\label{tab:bcp-dhcp-baseline}
\resizebox{\columnwidth}{!}{%
\begin{threeparttable}[t!]
\renewcommand{\arraystretch}{1.1}
\begin{tabular}{@{}|c|c|ccc|ccc|c|@{}}
\hline
\multirow{2}{*}{\textbf{Site}} & \multirow{2}{*}{\textbf{Method}} & \multicolumn{3}{c|}{\textbf{Agreement Rate (\%)}} & \multicolumn{3}{c|}{\textbf{Angular Error (°)\tnote{*}}} & \multirow{2}{*}{\textbf{$\Delta$AFD}} \\ \cline{3-8}
 &  & \multicolumn{1}{c|}{\textbf{1-F}} & \multicolumn{1}{c|}{\textbf{2-F}} & \textbf{3-F} & \multicolumn{1}{c|}{\textbf{1-F}} & \multicolumn{1}{c|}{\textbf{2-F}} & \textbf{3-F} &  \\ \hline
\multirow{2}{*}{dHCP} & DL & \multicolumn{1}{c|}{87.03} & \multicolumn{1}{c|}{23.78} & \enskip{}8.40 & \multicolumn{1}{c|}{8.95} & \multicolumn{1}{c|}{16.82} & 29.26 & 0.128 \\ \cline{2-9} 
 & GS & \multicolumn{1}{c|}{95.30} & \multicolumn{1}{c|}{57.10} & 55.28 & \multicolumn{1}{c|}{3.39} & \multicolumn{1}{c|}{\enskip{}7.64} & 20.79 & 0.050 \\ \hline
\multirow{2}{*}{BCP} & DL & \multicolumn{1}{c|}{81.17} & \multicolumn{1}{c|}{29.14} & 11.32 & \multicolumn{1}{c|}{9.02} & \multicolumn{1}{c|}{16.61} & 35.26 & 0.257 \\ \cline{2-9} 
 & GS & \multicolumn{1}{c|}{89.21} & \multicolumn{1}{c|}{52.67} & 46.19 & \multicolumn{1}{c|}{4.19} & \multicolumn{1}{c|}{\enskip{}9.38} & 24.34 & 0.075 \\ \hline
\end{tabular}%
\begin{tablenotes}
    \item[*] All AE values are computed among fibers with GT-matched peak predictions.
\end{tablenotes}%
\end{threeparttable}%
}
\end{table}

\begin{figure}[t!]
    \centering
    \includegraphics[width=\linewidth]{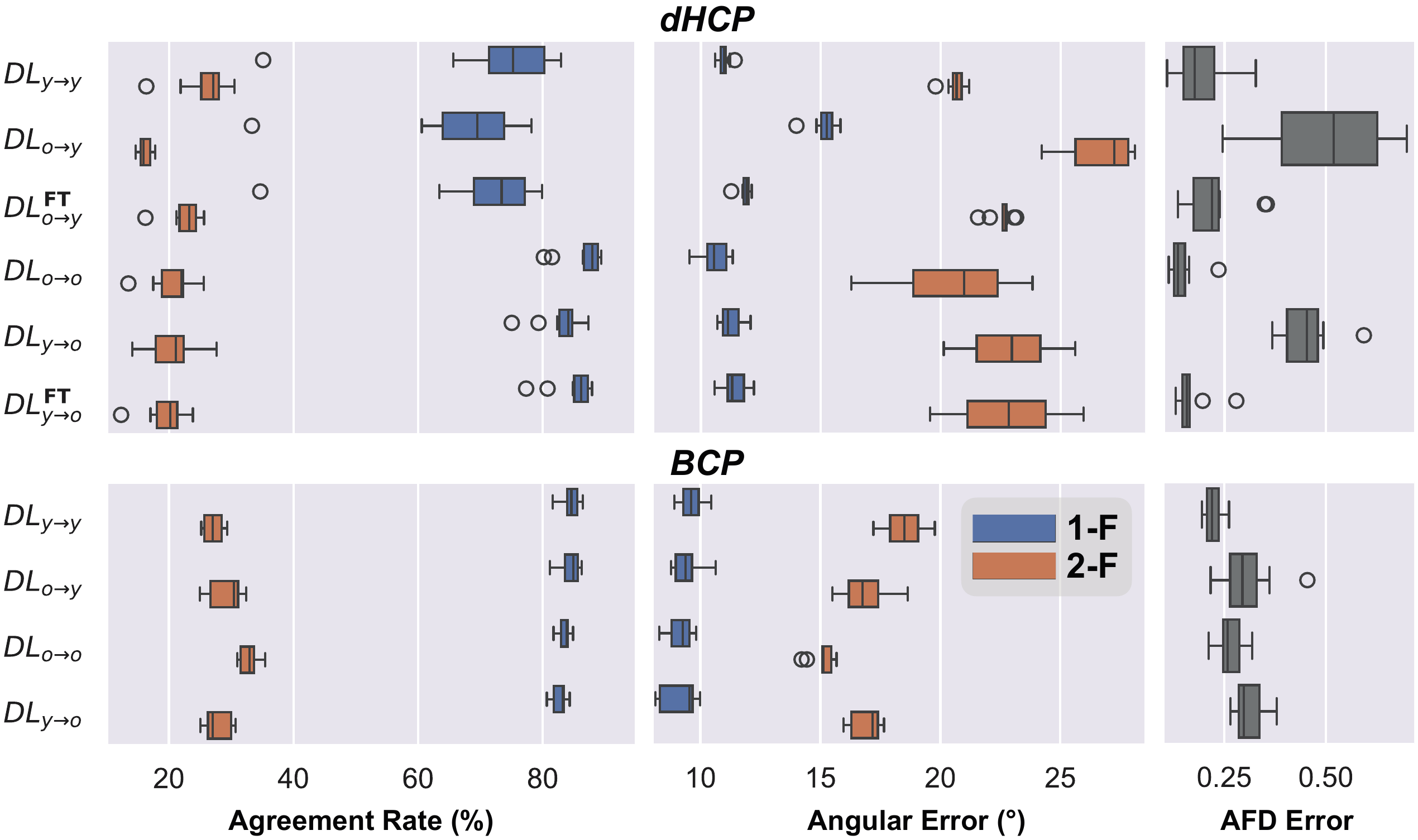}
\caption{
Comparative intra-site performance of DL models across age-specific training in \textbf{dHCP (top)} and \textbf{BCP (bottom)}, showing AR and AE under 1/2-F configurations alongside the AFD Error. $\text{DL}_{\text{a}\shortrightarrow\text{b}}$ denotes models trained on ``a'' and tested on ``b''; further fine-tuned on ``b'' when followed by $^{\mathsf{FT}}$.
}
    \label{fig:dhcp-bcp-intra}
\end{figure}

Moving to age-specific comparisons, we observed different patterns between younger and older age groups (denoted as ``y'' and ``o'', respectively), for dHCP and BCP datasets. \Cref{fig:dhcp-bcp-intra} illustrates these differences, revealing that age-related effects in BCP are less marked compared to dHCP. For instance, the difference in single-fiber ARs and AFD between DL$_{\text{y}\shortrightarrow\text{y}}$ and DL$_{\text{o}\shortrightarrow\text{o}}$   is higher within dHCP than BCP, approximately 14\% and 21\% for AFD error, respectively. 

\begin{figure}[t!]
    \centering
    \includegraphics[width=\linewidth]{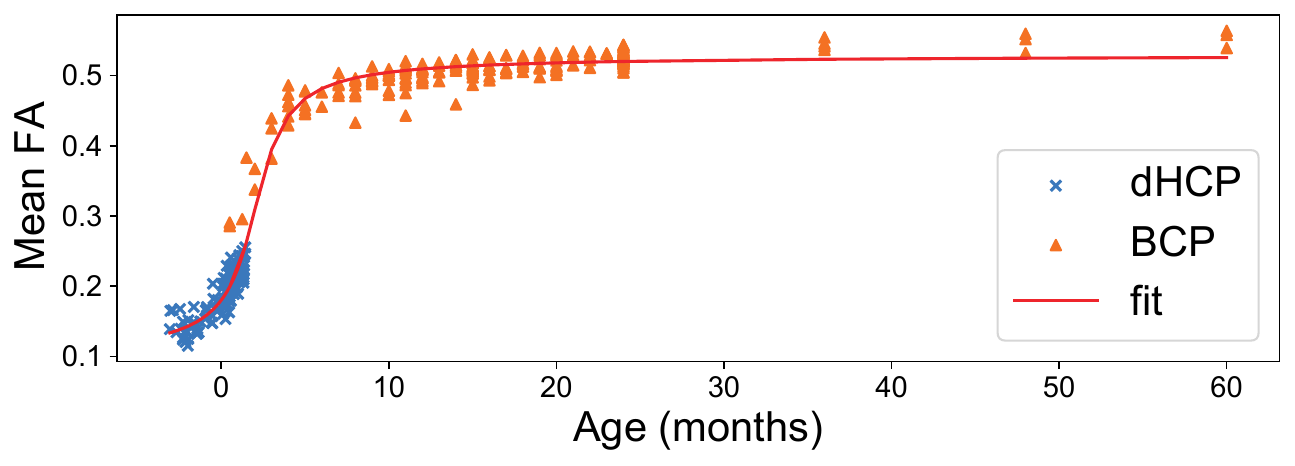}
\caption{
Mean FA within white matter area by postnatal age, with dHCP (\textcolor{MyBlue}{×}) shifted from post-menstrual to postnatal age and BCP (\textcolor{MyOrange}{$\blacktriangle$}), modeled with an $\arctan$ growth fit curve (red).
}
    \label{fig:age-fa}
\end{figure}

Within each age group, we see similar stability in BCP compared to dHCP when training on young and testing on old subjects or vice versa. This consistency could be due to rapid development and white matter changes in the first months of life, as opposed to slower white matter development in later periods \cite{yu_differential_2020}. To validate this hypothesis, we explored the average white matter FA in our cohorts (\Cref{fig:age-fa}), which indeed shows a significant increase in dHCP but a plateau in the BCP cohort. In general, AE seems to be less prone to age effect, except when training on older dHCP subjects and testing on younger ones (DL$_{\text{y}\shortrightarrow\text{y}}$), where there is also a more pronounced decline in AR.
Finally, fine-tuning on five dHCP subjects consistently reduces error rates, especially for AFD error.

\subsection{Inter-Site Experiments}

\begin{figure}[t!]
    \centering
    \includegraphics[width=\linewidth]{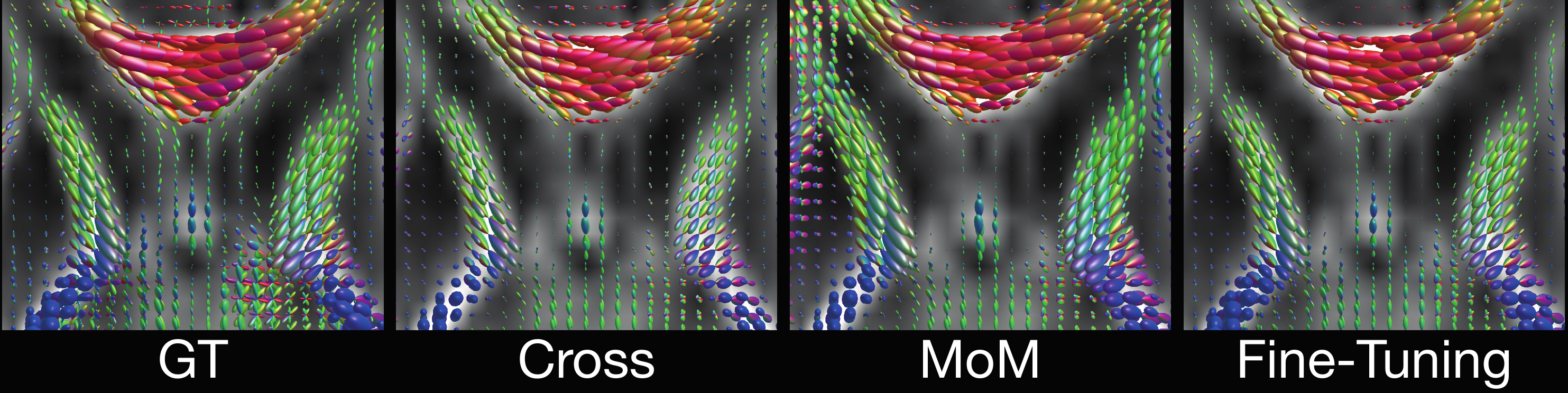}
    \caption{Qualitative comparison between GT and cross-site estimated FODs on dHCP, visualized on FA map.   }
    \label{fig:qualitative}
\end{figure}

\begin{figure}[t!]
    \centering
    \includegraphics[width=1\linewidth]{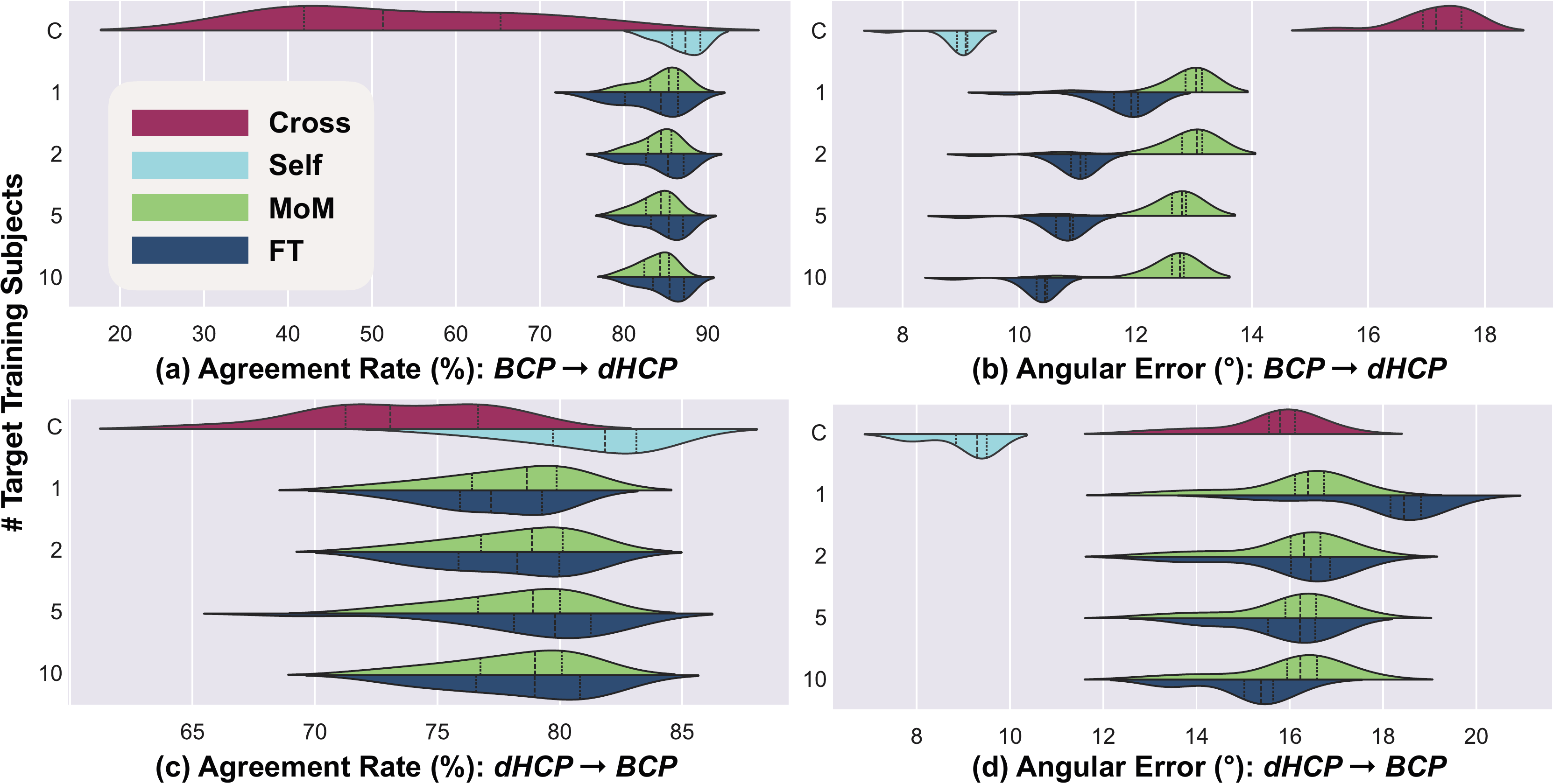}
    \caption{
    Inter-site performance of BCP-trained models tested on dHCP \textsf{(a, b)}, and dHCP-trained models tested on BCP \textsf{(c, d)}, comparing MoM and FT methods using varying subjects (1, 2, 5, 10) from the target domain under single-fiber configuration, with cross-testing and self-testing serving as lower- and upper-performance bounds (``\textsf{C}ontrol''), respectively.
    }
    \label{fig:res}
\end{figure}

Given the high number of domain shifts (scanner, protocol, age) and the low agreement of GS/DL of multiple fiber depictions (\Cref{tab:bcp-dhcp-baseline}), we compare the cross-site results on single fiber populations only. Inter-site performance, depicted in \Cref{fig:res}, shows the capability of the DL model to generalize across different datasets (as shown in \Cref{fig:qualitative}). 

AR when tested on dHCP (as shown in \Cref{fig:res} (a)) displays a marked increase with fine-tuning on one and two subjects. However, increments from 2 to 5 and 5 to 10 subjects offer only marginal gains in AR.
Fine-tuning exhibits an advantage over MoM particularly when transitioning from BCP to dHCP.  
In both directions,  AE presents a slight improvement when more target subjects are incorporated into fine-tuning the model, although the extent of this improvement is not much pronounced (2--3\textdegree). Moreover, the MoM harmonization is less sensitive to the number of subjects used, both for AE and AR. As for the ablation study, where the model was trained solely on 10 target dataset subjects, revealed notably reduced performance.

In summary, the inter-site experiments show that refining the DL-based fiber estimation pipeline using few target domain subjects in fine-tuning or MoM outperforms direct cross-testing and can make the accuracy closer to direct testing in some cases such as AR in dHCP testing shown in \Cref{fig:res} (a). This improvement is significantly more visible for transfers from BCP to dHCP than vice versa, suggesting some distinct dynamics at play when adapting a model from a dataset with pronounced age-related shifts (dHCP) to one with more gradual changes (BCP), and less so in the reverse direction.

\section{Conclusion}
\label{sec:conclusion}

This work has demonstrated that even a small number of target data samples can be instrumental in overcoming domain shifts encountered in white matter fiber estimation with deep learning. Through the application of fine-tuning and to a lesser extent the MoM harmonization strategies, models have shown improved performance in estimating the FODs in developing brains in both cross-age and cross-site and age settings.
Moreover, we observed that the lower variations in the microstructural development of babies compared to newborns, have a direct influence on the performance of the DL models in the cross-age experiments. Such findings highlight the importance of tailoring DL models to account for the unique developmental stages of pediatric populations.

\section{Compliance with Ethical Standards}
\label{sec:ethics}

This retrospective research study used open-source human subject data from the Developing Human Connectome Project and the Baby Connectome Project, respectively, where ethical approval was \textbf{not} required per the data licenses.

\section{Acknowledgments}
\label{sec:acknowledgments}

We acknowledge the CIBM Center for Biomedical Imaging, a Swiss research center of excellence founded and supported by CHUV, UNIL, EPFL, UNIGE, HUG, and the Leenaards and Jeantet Foundations. This research was partly funded by the Swiss National Science Foundation (grants 182602, 215641); 
also by the National Institute of Neurological Disorders and Stroke and the Eunice Kennedy Shriver National Institute of Child Health and Human Development of the US National Institutes of Health (award numbers R01NS106030, R01NS128281, R01HD110772).
We thank Hakim Ouaalam for preprocessing and parcellating BCP T2-weighted images.


\printbibliography[title=References]

\end{document}